\documentclass[twocolumn,showpacs,showkeys,preprintnumbers,amsmath,amssymb,aps]{revtex4-1}

\usepackage{amsfonts}
\usepackage{amssymb}
\usepackage{amsmath}
\usepackage{graphicx}
\usepackage{float}
\usepackage{color}

\begin{document}

\title{Topological Insulators in Ternary Compounds with a Honeycomb Lattice}

\author{Hai-Jun Zhang$^{1}$,  Stanislav Chadov$^{2}$, Lukas M\"{u}chler$^{2}$, Binghai Yan$^{1}$, Xiao-Liang Qi$^{1}$, J\"{u}rgen K\"{u}bler$^{3}$, Shou-Cheng Zhang$^{1}$, Claudia Felser$^{1,2}$}

\affiliation{
  $^1$Department of Physics, McCullough Building, Stanford University,
  Stanford, California 94305-404531\\
  $^2$Institut f\"ur Anorganische Chemie und Analytische
  Chemie, Johannes Gutenberg - Universtit\"{a}t,  55099 Mainz, Germany,\\
  $^3$Institut f\"{u}r Festk\"{o}rperphysik, Technische Universit\"{a}t Darmstadt, 64289 Darmstadt, Germany
}

\email{felser@uni-mainz.de}

\date{\today}



\begin{abstract}
  
One of the most exciting subjects in solid state physics is a
 single layer of  graphite which exhibits a variety of
 unconventional novel properties. The key feature of its
electronic structure are linear dispersive bands which cross in a
single point at the Fermi energy. This so-called Dirac cone is
closely  related  to  the  surface  states of  the  recently  discovered
topological insulators. The ternary compounds,
such as LiAuSe and KHgSb with a honeycomb structure of their Au-Se and Hg-Sb
layers feature   band inversion very  similar to HgTe  which is a
strong precondition for existence of the topological surface states.
In contrast to graphene with two Dirac cones at $K$ and $K'$
points, these materials exhibit the surface states formed by only a single Dirac cone at the $\Gamma$
point together with the small direct band gap opened by a strong spin-orbit
coupling (SOC) in the bulk. These materials are  centro-symmetric,
therefore, it is possible  to  determine the  parity of  their
wave functions, and hence, their topological character.
Surprisingly, the compound KHgSb with the strong SOC is topologically trivial,
whereas LiAuSe is found to be a topological non-trivial insulator.

\end{abstract}

\maketitle

The search for new materials with inverted band structure  provides
the basis for the Quantum Spin Hall effect (QSH)\cite{Bernevig06,Koenig07,Fu07,FKM07,Dai08,Hsieh08,Xia09,Zhang09,Chen09,CQK+10,LWX+10,Qi2010,Moo10}.
This new exciting field started with the prediction and experimental
observation of the QSH in quantum wells in two-dimensional topological insulator of the binary semiconductor HgTe~\cite{Bernevig06,Koenig07}.
In a series of single crystals, such as Bi$_2$Se$_3$, three-dimensional
topological insulting behavior was observed in
topological surface states appearing as Dirac cones~\cite{Xia09,Zhang09,Chen09}.
Later on, the manifold of Heusler semiconductors with 18 valence
electrons and a similar band inversion was proposed ~\cite{CQK+10,LWX+10}.
Since this proposed class of materials is extremely rich, it provides
much wider flexibility in design  by tuning the band gap size and the spin-orbit coupling
(SOC) magnitude. In addition, the multi-functionality allows the
incorporation of new properties such as superconductivity or magnetism~\cite{CQK+10}.

The structure of the XYZ Heusler compounds
can be simply viewed as ``stuffed'' YZ-zinc-blende. Depending on the stuffing
element X Heuslers are semiconducting or semi-metallic~\cite{Kandpal06}.
Materials like ScPtBi are topologically similar to
HgTe: the inversion of the conduction and valence bands occurs due
to small electronegativity differences. Since HgTe and ScPtBi are both 2D topological
insulators, the QSH is also expected in the corresponding quantum
wells, as e.\,g. ScPtSb/ScPtBi, in full analogy to CdTe/HgTe. We emphasize that the  check of  parity at the time reversal points, as a sufficient condition for
their topological character, is not possible here because of the absence of  inversion
symmetry~\cite{Fu07}.

Heusler compounds are similar to a  stuffed diamond, correspondingly, it should be possible to find the
``high Z'' equivalent of graphene in a graphite-like
structure with 18 valence electrons and with inverted bands. In this
structure type with a lower symmetry
compared to diamond three-dimensional topological behavior is
realizable. Indeed, the honeycombs KZnP and KHgSb,
crystallizing in so-called AlB$_2, $ Ni$_2$In
or ZrBeSi structure types, exhibit  band ordering similar to the cubic CdTe and
HgTe. However, in contrast to graphene, these compounds have a
strong SOC which leads to a finite band gap at the $\Gamma$ point~\cite{KM05a,Bernevig06a}. The
determination of the wave function parity at the time-reversal  points for these materials~\cite{Fu07}
is now possible, since they are centro-symmetric.

\begin{figure}
\centering
\begin{minipage}[b]{0.4\linewidth}
 (a)\includegraphics[width=1.0\textwidth]{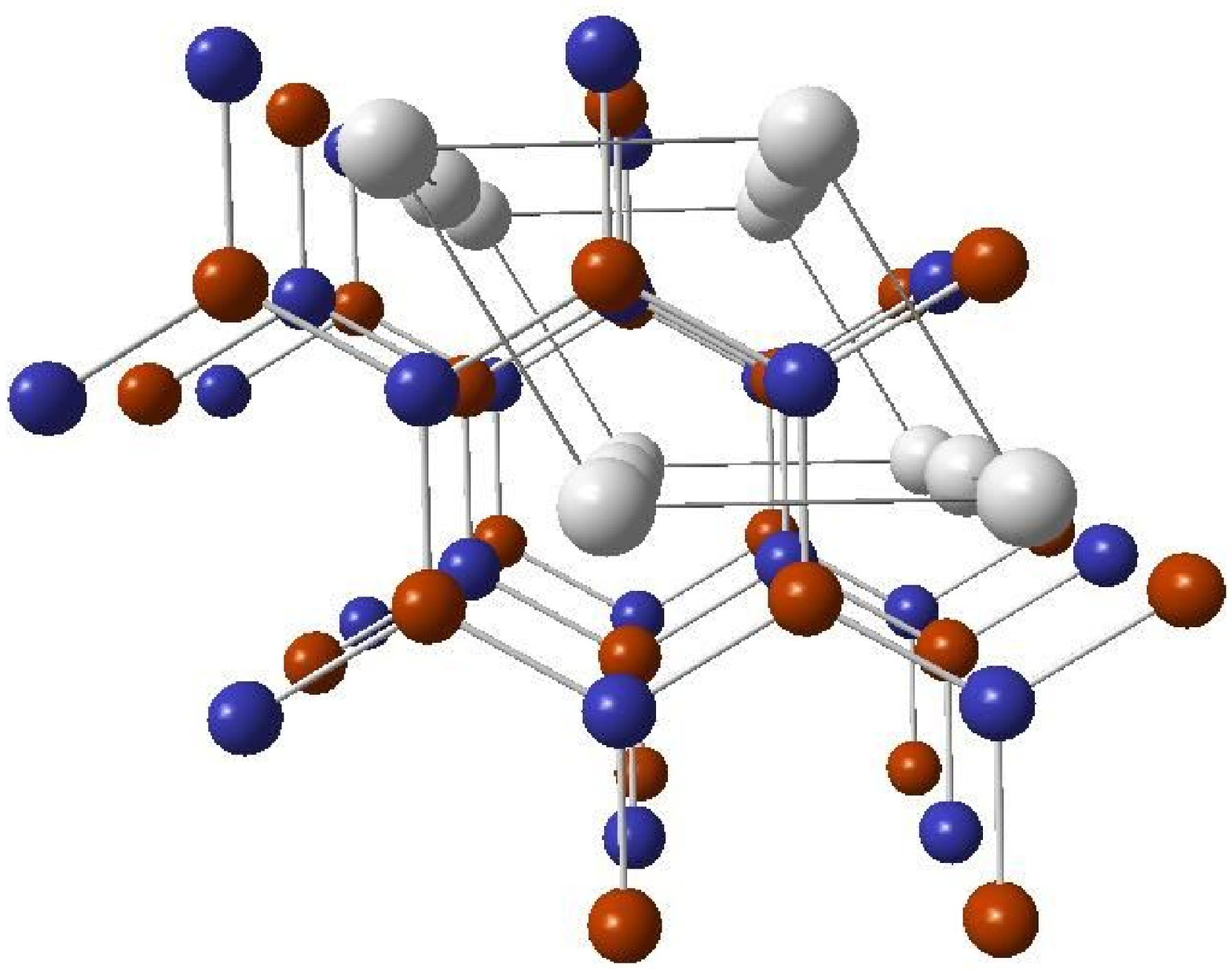}
\end{minipage}~~
\begin{minipage}[b]{0.4\linewidth}
 (b)\includegraphics[width=1.0\textwidth]{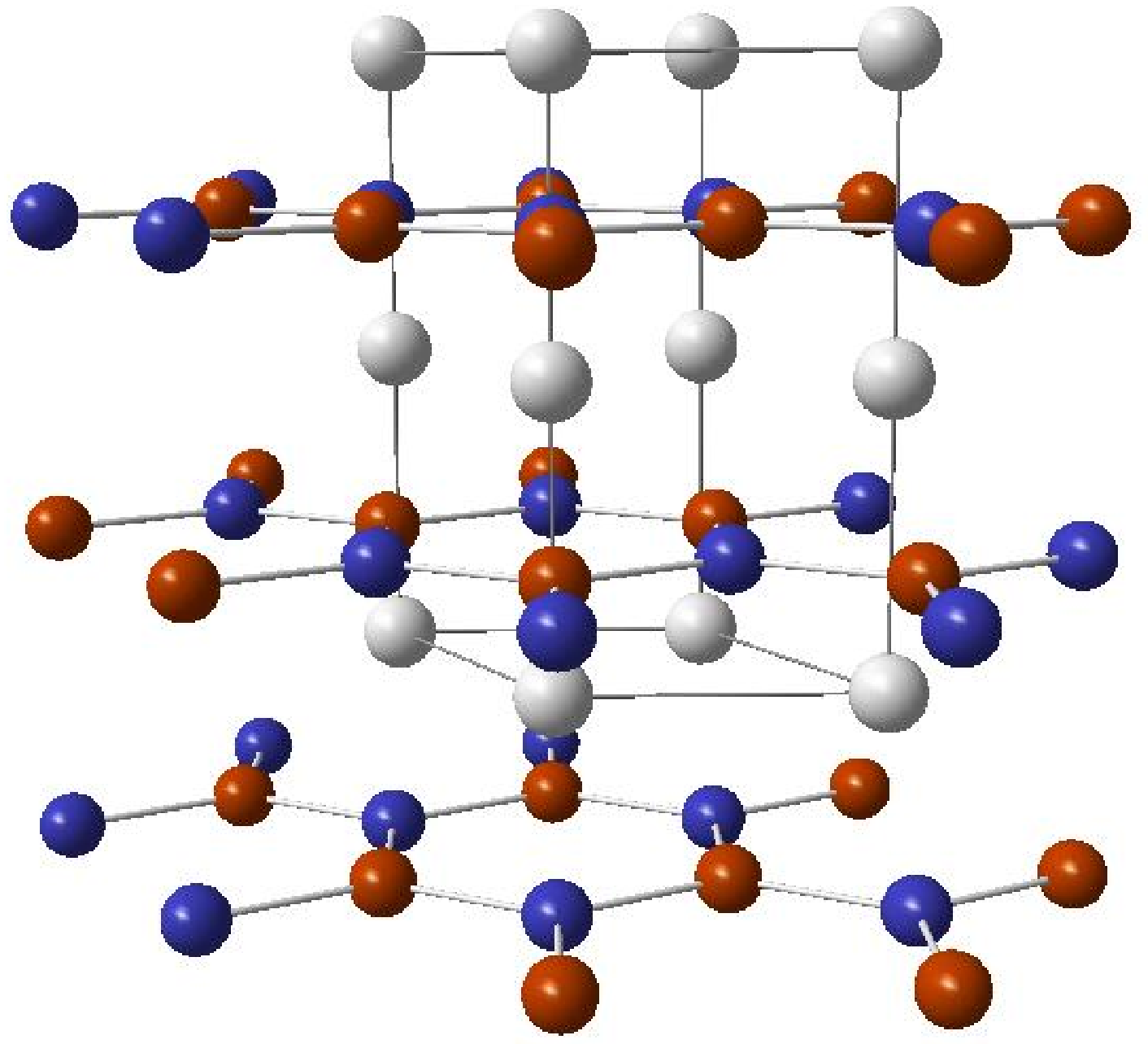}
\end{minipage}
\caption{ (color online)
 Structure of KZnP, KHgSb, KCuSe and LiAuTe (ZrBeSi structure type) (a) shown in the
 $a$-$b$ plane  and (b) along the $c$ axis. Thus, the ternary
 semiconductor (with 18 valence electrons) can be viewed
 as stuffed relative of graphite with the honeycomb lattice
 consisting of alternating late transition metals (Ag, Au, Zn, Hg)
and main group elements (Se, Te, P, As, Sb). }
\label{FIG:str-graphite}
\end{figure}

In detail,  KHgSb, for instance,  can be presented as a stuffed graphene
in the following way: the electropositve K$^+$ is stuffed in a honeycomb
lattice of [HgSb]$^-$~\cite{CFS+08}.  The ZrBeSi structure with additional stuffing
electropositive main group elements is illustrated in Figure~\ref{FIG:str-graphite}.
One sees that the main group and the transition elements alternate
within the same layer and between the layers, therefore the primitive
unit cell consists of two formula units.

The number of ternary compounds crystallizing in this structure
 is nearly as large as in the Heusler family, however due to a weaker mean
hybridization ($sp^2$ versus $sp^3$)  many of the 18 valence electron compounds exhibit
substantially  smaller gaps or are simply  metallic. As we show in the following, a certain number of  semiconducting materials
of this family with ordinary and inverted band
structures can be found.  In contrast to HgTe and ScPtBi,
 such materials with inverted structure exhibit real band gaps
due to their higher structural anisotropy. For this reason the
honeycomb compounds are three dimensional (3D), rather than 2D
topological insulators.  The 3D topological insulators exhibit
an insulating energy gap in the bulk and gapless states on the
surface  protected by time-reversal symmetry~\cite{FKM07}. These surface
states exist only if there is an odd number of massless Dirac cones,
with a single cone in the simplest case. The oddness is provided by the
$Z_2$ topological invariant of the bulk~\cite{KM05a,KM05b}, thus any time-reversal invariant
perturbation cannot open an insulating gap at the Dirac point on the
surface.

The bulk band gap is defined by interplay of the lattice
constant, spin-orbit coupling and the difference in
electronegativities of the honeycomb sublattice constituents.
In analogy to a ternary cubic semiconductors, new topological insulators
can be found among the heavy 8/18 valence electron relatives of graphene, i.\,e. in a
graphite XYZ structure type. Suitable atomic
combinations are:  (i) X\,=\,Li, Na, K, Rb, Cs; Y\,=\,Zn, Cd, Hg and Z\,=\,P,
As, Sb, Bi, or (ii) X\,=\,K, Rb, Cs; Y\,=\,Ag, Au and Z\,=\,Se, Te, or (iii) X\,=\,rare earth,
Y\,=\,Ni, Pd, Pt and Z\,=\,P, As, Sb, Bi.  Some of these combinations
are not yet synthesized, some crystallize in different forms
such as variants of the Cu$_2$Sb structure type (they will be the
subject of another publication) and some, especially the rare earth containing
compounds, are metallic. The known examples are  KZnAs, KZnSb, KZnP, KHgAs,
KHgSb, RbZnP, RbZnAs, RbZnSb, NaAuTe, KCuSe, KCuTe, KAuTe, and RbAuTe.
Compounds such as LiAuSe, LiAuTe, CsAuTe, KHgBi, and CsHgBi are likely to be synthesized
and we consider them in the present study as well.
As mentioned above, the electronic structure of these materials is similar to their cubic analogues.
The alkaline ions Li$^+$, Na$^+$, K$^+$, Rb$^+$, and Cs$^+$ ``stuff'' the
graphite type YZ planar sublattice.  Since these 18-electron compounds form such closed-shell
structures, they are all non-magnetic and semiconducting.
The difference is that in the case of binary semiconductors or C$_{1b}$ Heuslers, the
bonds within the YZ tetrahedrons are of $sp^{3}$ or $sd^{3}$ type,
whereas in planar graphite-type layers the $\sigma$-type bonding
occurs between the $sp^{2}$ or $sd^{2}$ orbitals. The remaining
$p$-orbitals provide the $\pi$-type bonding interaction, similar to graphite.

The search for the topological character of the proposed materials is
based on {\it ab-initio} calculation of the electronic
structure (for details see the supplemental).
An important peculiarity of hexagonal systems is emphasized here: their unit cell
consists of two formula units which  results in a
doubling of the corresponding bands.  More details are given in the 
supplemental (Fig.~1)  where it is seen that for KZnP (KHgSb)
at the $\Gamma$ point the doubled  $s$-bands are split by about
10\,meV. Thus for the weak coupling  of the nearest honeycomb planes these
doubled bands can become nearly indistinguishable.
In the following we show that such doubling of the unit
cell leads to a change of topology.  Fortunately, the
 centro-symmetric  space group (194) of the honeycomb-type compounds allows
to make use of the parity eigenvalues~\cite{Fu07, Zhang09}.
All relevant  properties, i.\,e. the wave function parity in
time-reversal symmetric $k$-points ($\Gamma(0,0,0), M(\pi,0,0), L(\pi,0,\pi),
A(0,0,\pi)$), the $Z_2$ invariant, the average nuclear charge
$\left<Z\right>$, the band gap width $E_{\rm g}$ and corresponding
lattice parameters are listed in Table~I for the compounds studied here.

\begin{table}[htb!]
\centering
\caption{Wave function parities at the time-reversal points, $Z_2$
  topological invariant~\cite{Fu07}, average nuclear charge $\left<Z\right>$, band gap $E_{\rm g}$ and optimized lattice parameters.}
\begin{tabular}{|c|c|c|c|c|c|c|c|c|c|c|}
\hline
 & $\Gamma$ & $M$ & $L$ & $A$ &
$Z_{2}$ & $\left<Z\right>$ & $E_{\rm g}$\,[meV]&$a$\,[au]&$c$/$a$\tabularnewline
\hline
\hline
LiAgSe & $+$ & $-$ & $-$ & $-$ & $1$ & 28 & 1&8.507& 1.512\tabularnewline
\hline
LiAgTe & $-$ & $-$ & $-$ & $-$ & $0$ & 34  & 45&8.979&1.526\tabularnewline
\hline
LiAuSe & $+$ & $-$ & $-$ & $-$ & $1$ & 38.(6)& 50&8.353&1.670\tabularnewline
\hline
LiAuTe & $+$ & $-$ & $-$ & $-$ & $1$ & 44.(6)  & -&8.818&1.663\tabularnewline
\hline
NaAgSe & $+$ & $-$ & $+$ & $+$ & $1$ & 30.(6)& 10&8.544&1.737\tabularnewline
\hline
NaAgTe & $+$ & $-$ & $+$ & $+$ & $1$ & 36.(6)& 3&9.031&1.719\tabularnewline
\hline
NaAuSe & $+$ & $-$ & $+$ & $+$ & $1$ & 41.(3)& 15&8.427&1.855\tabularnewline
\hline
NaAuTe & $+$ & $-$ & $+$ & $+$ & $1$ & 47.(3)& 30&8.894&1.823\tabularnewline
\hline
KAgSe & $-$ & $-$ & $+$ & $+$ & $0$ & 33.(3) & 15&8.836&1.989\tabularnewline
\hline
KAgTe & $-$ & $-$ & $+$ & $+$ & $0$ & 39.(3)& 230&9.193&1.958\tabularnewline
\hline
KAuSe & $-$ & $-$ & $+$ & $+$ & $0$ & 44& 20&8.742&2.077\tabularnewline
\hline
KAuTe & $+$ & $-$ & $+$ & $+$ & $1$ & 50& -&8.780&2.097\tabularnewline
\hline
\hline
LiZnAs & $-$ & $-$ & $-$ & $-$ & $0$ & 22& 280&7.881&1.791\tabularnewline
\hline
LiZnSb & $-$ & $-$ & $-$ & $-$ & $0$ & 28 & 10&14.121&1.762\tabularnewline
\hline
LiHgAs & $+$ & $-$ & $-$ & $-$ & $1$ & 38.(6)& -&8.542&1.686\tabularnewline
\hline
LiHgSb & $+$ & $-$ & $-$ & $-$ & $1$ & 44.(6) & -&9.070&1.665\tabularnewline
\hline
NaZnAs & $-$ & $-$ & $+$ & $+$ & $0$ & 24.(6)& 80&7.952&2.083\tabularnewline
\hline
NaZnSb & $-$ & $-$ & $+$ & $+$ & $0$ & 30.(6)& 230&8.539&2.007\tabularnewline
\hline
NaHgAs & $-$ & $-$ & $+$ & $+$ & $0$ & 41.(3)& -&8.615&1.897\tabularnewline
\hline
NaHgSb & $-$ & $-$ & $+$ & $+$ & $0$ & 47.(3) & 140&9.148&1.845\tabularnewline
\hline
KZnAs & $-$ & $-$ & $+$ & $+$ & $0$ & 27.(3)& 160&7.993&2.419\tabularnewline
\hline
KZnSb & $-$ & $-$ & $+$ & $+$ & $0$ & 33.(3)& 130&8.654&2.349\tabularnewline
\hline
KHgAs & $-$ & $-$ & $+$ & $+$ & $0$ & 44 & 80&8.515&2.214\tabularnewline
\hline
KHgSb & $-$ & $-$ & $+$ & $+$ & $0$ & 50 & 250&9.040&2.140\tabularnewline
\hline
\end{tabular}
\end{table}

From previous studies on binary and ternary cubic
semiconductors~\cite{CQK+10} it follows that  it is more probable to find the  $Z_2=1$
topological insulator among heavier compounds (with stronger SOC). Indeed, their bands
splitting scales roughly with the average nuclear charge ${\left<Z\right>=1/N\sum_{i=1}^{N}Z_i}$ where $N=2$ for
binaries and $N=3$ for ternaries. This   parameter sorts cubic
systems almost along a straight line~\cite{CQK+10}. However Table~I clearly illustrates  that this does not
hold  for the semiconductors of the ZrBeSi structure type.
Indeed, only  Lithium compounds  within the Zn and Hg group
show the expected trend.  The compounds with Zn  are
topologically trivial  whereas those with Hg are non-trivial insulators.
Compounds with heavier alkaline metals (Na, K) are all
 trivial independently of whether or not they contain Zn or Hg.
Among  the Ag- and Au-containing compounds  more non-trivial systems are found,
however the correlation between band  inversion and   $\left<Z\right>$, as in the cubic semiconductors, is
 absent. For example, the topological insulator LiAgSe corresponds
 to $\left<Z\right>=28$,   whereas for the topologically trivial KHgSb system  $\left<Z\right>=50$.

\begin{figure}
\centering
  \includegraphics[angle=270,width=0.85\linewidth,clip]{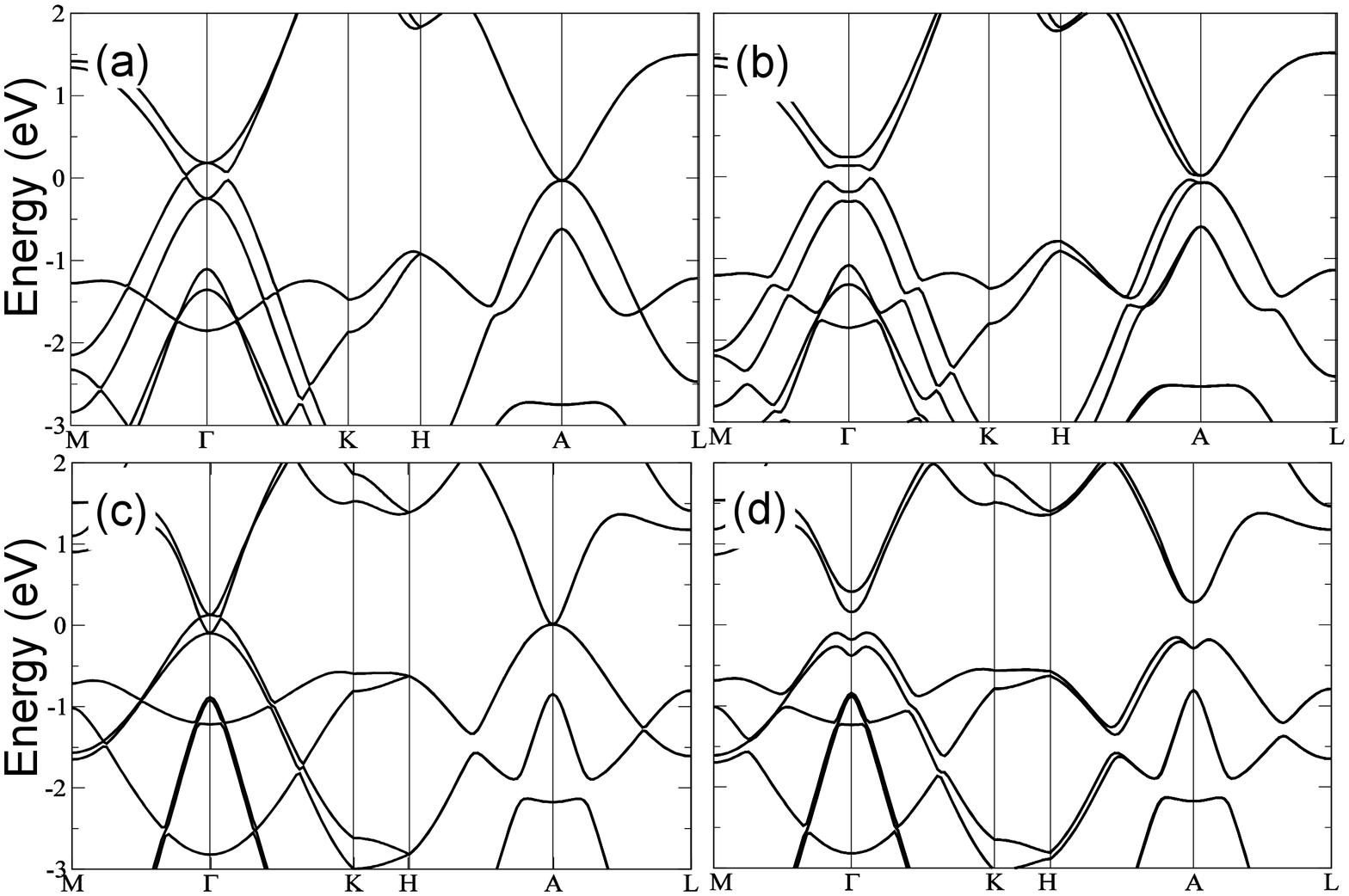}\\[1ex]
\caption{ Band structure of LiAuSe (a, b) and KHgSb (c, d) calculated
  without (a, c) and with (b, d) SOC.}
\label{FIG:bns-transformations}
\end{figure}

 In Figure~\ref{FIG:bns-transformations} the band structure of
 LiAuSe (upper panel) is compared with KHgSb (lower panel) calculated at 
 time-reversal symmetric points  with (right) and without (left) SOC.
It follows that both compounds are semi-metals with degeneracies
 at the $\Gamma$ and $A$ symmetry points if SOC is omitted.
The degenerate $p_{\rm x}$ and $p_{\rm y}$ states  mediate $\sigma$-type Sb/Se and
Hg/Au bonding.   The lower-lying Hg/Au bands
 are of $s$ type, similar to HgTe.
Inclusion of  SOC opens a band gap at the Fermi energy
at the  $\Gamma$ and $A$ points leading to the  typical dips in the band structure 
 for both compounds~\cite{Kli10}. The resulting parities for both materials listed in Table~I
match  only at the $M$ point, whereas  at $\Gamma$, $A$, and $L$ they  differ,
leading to a trivial state in KHgSb and  topological insulator in LiAuSe.
\begin{figure}
\centering
\begin{minipage}[t]{0.4\linewidth}
  \includegraphics[angle=270,width=1.0\linewidth,clip]{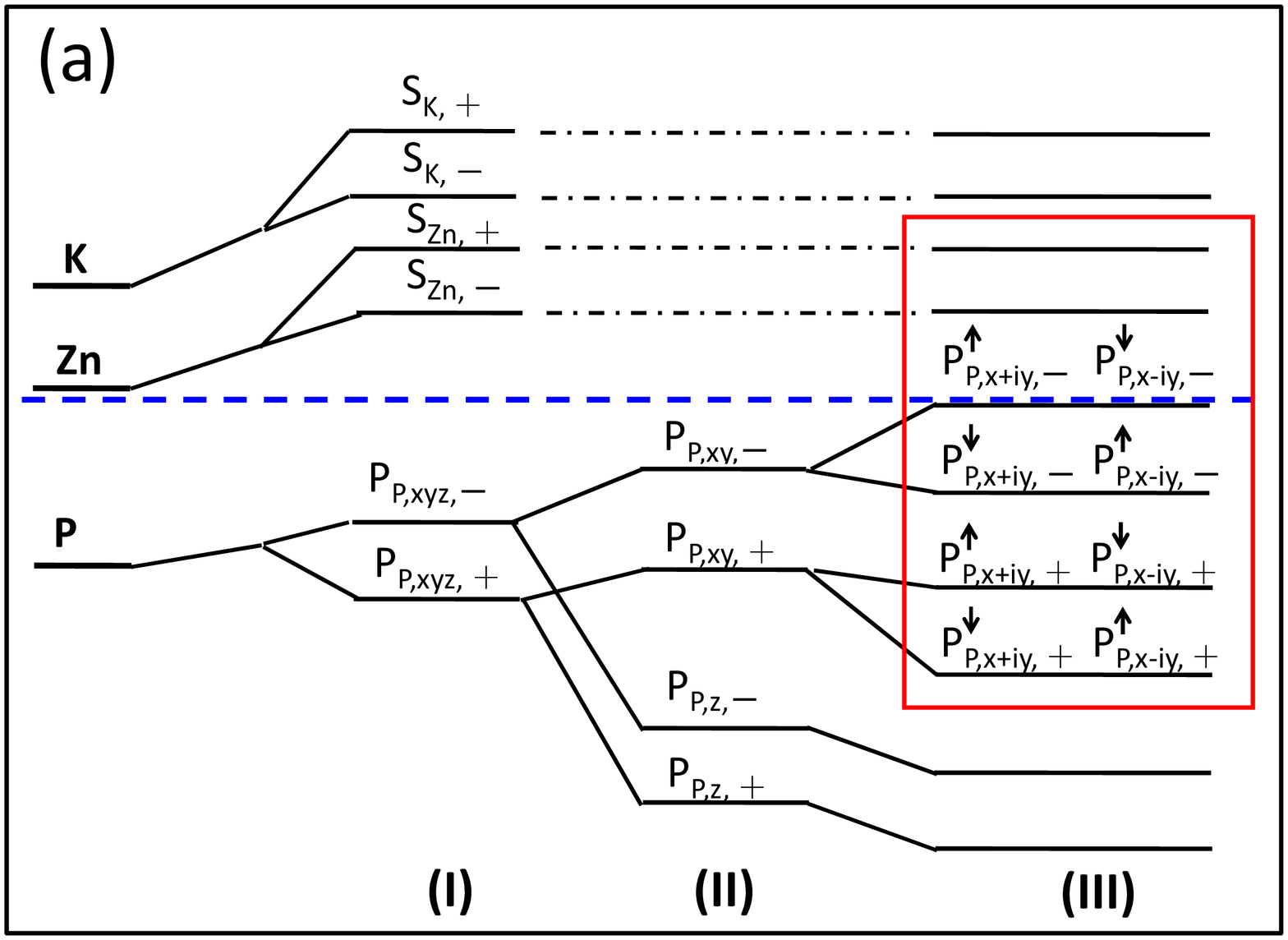}
\end{minipage}~
\begin{minipage}[t]{0.4\linewidth}
  \includegraphics[angle=270,width=1.0\linewidth,clip]{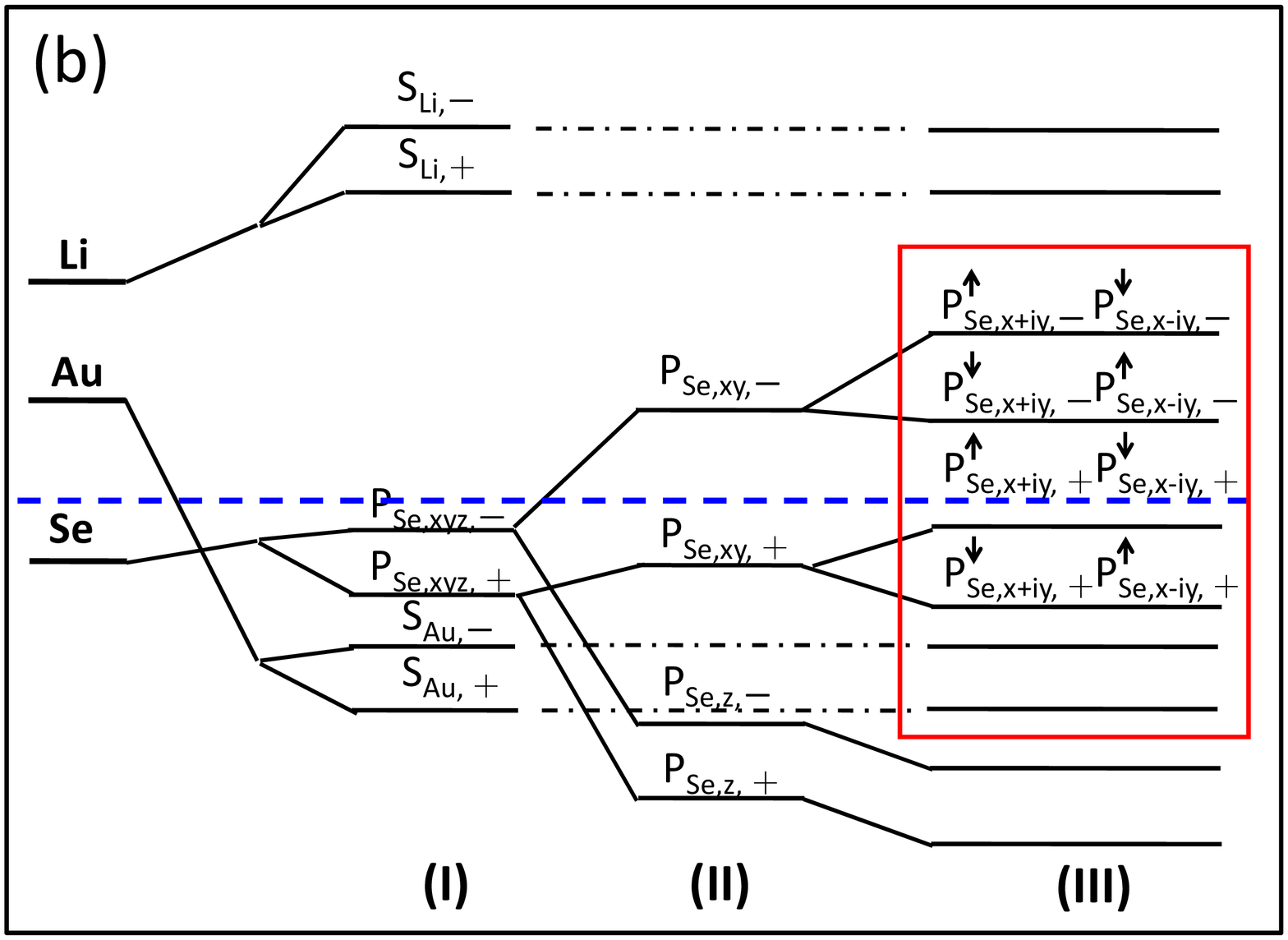}
\end{minipage}

\begin{minipage}[t]{0.4\linewidth}
  \includegraphics[angle=270,width=1.0\linewidth,clip]{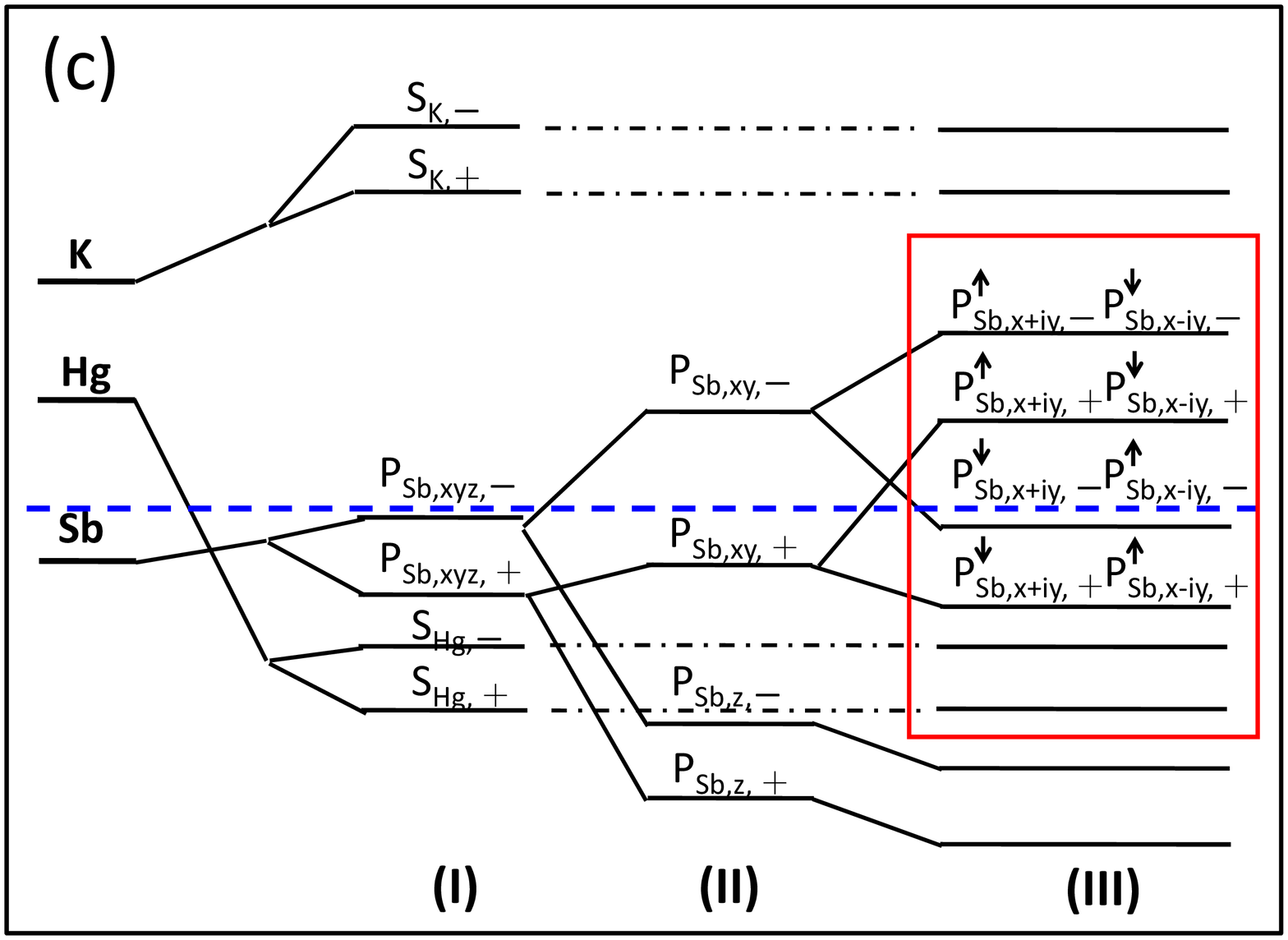}
\end{minipage}~
\begin{minipage}[t]{0.4\linewidth}
  \includegraphics[angle=270,width=1.0\linewidth]{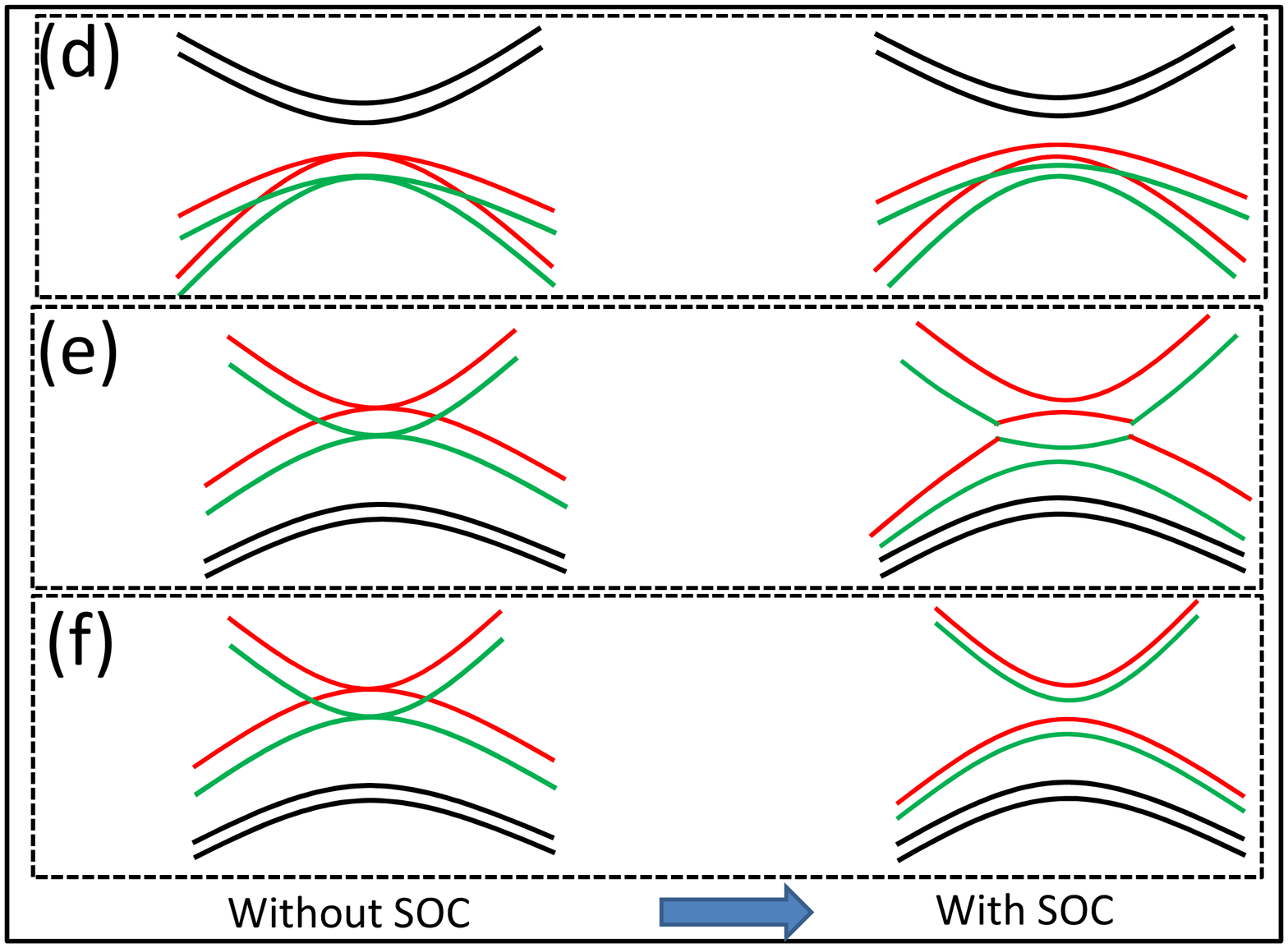}
\end{minipage}
\caption{(Color online) The structure of the atomic orbitals.
(a) Evolution of atomic orbitals at the $\Gamma$ point in KZnP system,
which has no inverted bands formed by Zn $s$ and P $p_{\rm x,y,z}$-orbitals.
Steps (I-III) represent the effect of
turning on chemical bonding (I), crystal-field splitting (II) and SOC (III).
The blue dashed line marks the Fermi energy.
Similar scheme for LiAuSe (b) and KHgSb (c),
(d), (e) and (f) represent the schematic band structure for KZnP, LiAuSe
and KHgSb near the $\Gamma$ point, respectively. The black curves mark the orbitals
with mainly Zn (or Au and Hg) $s$-character, red and green curves are the rest of light
and heavy hole-like bands.}
\label{FIG:bns-transformations-scheme}
\end{figure}

To understand the mechanisms of the band inversion and  parity
 change in details one can track the band structure evolution near the $\Gamma$ point by
 starting from the simple atomic energy levels and subsequently introducing
  chemical bonding (step I), crystal field (II), and SOC (III). The resulting changes are
  schematically shown in Fig.~\ref{FIG:bns-transformations-scheme}.  As
  an example we took the KZnP system. The low-lying $3s$-orbital of
  P($3s^23p^3$) can be neglected and the consideration can be restricted to the $s$-orbitals of
  K($4s$) and Zn($3d^{10}4s^2$) and the $p_{x,y,z}$ orbitals of
  P($3s^23p^3$).  In step I the chemical bonding is
  introduced. It is convenient to make use of the inversion symmetry
  and recombine the orbitals  according to their parities. K and Zn give
  two even and two odd $s$-orbitals $\left|S_{\,\rm K,\pm}\right>$, $\left|S_{\,\rm Zn,\pm}\right>$;  P gives three odd
  and three even $p$-orbitals $\left|P_{\,\rm P,{xyz},\pm}\right>$, where ``$\pm$''
  are the parity labels. In step II the crystal field is switched
  on which splits the $\left|P_{\,\rm P,{xyz},\pm}\right>$ into
  $\left|P_{\,\rm P,{xy},\pm}\right>$ and $\left|P_{\,\rm P,z,\pm}\right>$ according
  to the hexagonal symmetry. In step III the SOC is turned on.
  This leads to a splitting between   $\left|P_{\,\rm P,{x+iy},\pm,\downarrow(\uparrow)}\right>$ and
  $\left|P_{\,\rm P,{z},\pm,\uparrow(\downarrow)}\right>$ orbitals. The Fermi level
  falls  into the middle of $\left|P_{\,\rm \!P,{x+iy},-,\downarrow}\right>$
  (or $\left|P_{\,\rm P,{x-iy},-,\uparrow}\right>$) and
  $\left|P_{\,\rm P,{x+iy},+,\uparrow}\right>$ (or
  $\left|P_{\,\rm P,{x-iy},+,\downarrow}\right>$) as shown in
  Figure~\ref{FIG:bns-transformations-scheme}\,(a).
 No inversion occurs between the $s$-orbital of K (or Zn) and
 $p$-orbital of P, similar to the CdTe case. However, this situation
 changes for the ${\rm LiAuSe}$ and KHgSb compounds. Due to the very delocalized
 character of Au (Hg) $d$-orbitals the $s$-orbital of Li (K)
 is pulled down below the $p$-orbital of P, which leads to  band
 inversion similar to the  HgTe case~\cite{delin2002}. The corresponding
 evolution for LiAuSe and KHgSb is shown in
 Figures~\ref{FIG:bns-transformations-scheme}\,(b) and (c). In addition,
 $\left|P_{\,\rm Sb,x+iy,-,\downarrow}\right>$ (or $\left|P_{\,\rm
   Sb,x-iy,-,\uparrow}\right>$) and $\left|P_{\,\rm
   Sb,x+iy,+,\uparrow}\right>$ (or $\left|P_{\,\rm
   Sb,x-iy,+,\downarrow}\right>$) are inverted because of the strong SOC
 in KHgSb. We emphasize that since there are two band inversions occurring in
 KHgSb it becomes topologically trivial. In contrast, LiAuSe exhibits  inversion
 only between  the $s$- and $p$-orbitals since its
 SOC is too weak to invert $\left|P_{\,\rm Sb,x+iy,-,\downarrow}\right>$
 (or $\left|P_{\,\rm Sb,x-iy,-,\uparrow}\right>$) and $\left|P_{\,\rm
   Sb,x+iy,+,\uparrow}\right>$ (or $\left|P_{\,\rm
   Sb,x-iy,+,\downarrow}\right>$). Thus we conclude that LiAuSe is
 topologically non-trivial.  The  corresponding  band structures around the $\Gamma$ point for KZnP,
 LiAuSe and KHgSb are shown schematically in
 Figures~\ref{FIG:bns-transformations-scheme}\,(d-f). It is seen that in KZnP  no band inversion occurs, in LiAuSe it occurs once and KHgSb  twice.

\begin{figure}
\centering
  \includegraphics[angle=270,width=0.9\linewidth,clip]{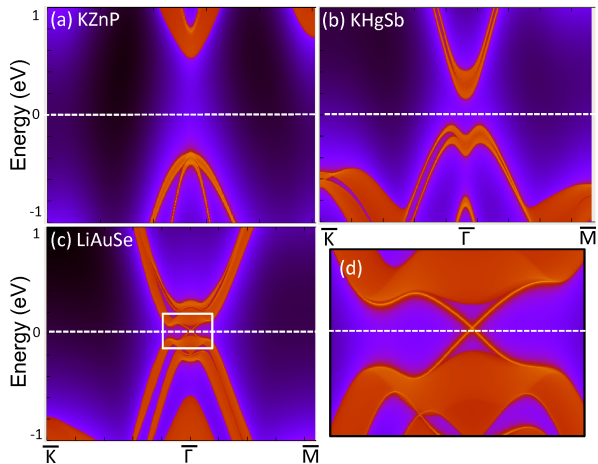}
\caption{ (color online) The band structure of the
semi-infinite (vacuum/solid-interface)  corresponding to (a)  trivial KZnP,
(b)  non-trivial  LiAuSe and (c) trivial KHgSb insulators, calculated along the $\bar{K}$-$\bar{\Gamma}$-$\bar{M}$
directions of the Brillouin zone. In contrast to KZnP and KHgSb, LiAuSe exhibits the surface
states seen in the bulk energy gap (zoomed in the inset (d)).}
\label{FIG:surface-states}
\end{figure}

For the final test of the topological/trivial character we directly calculate the surface
states of the proposed systems. The electronic structure of the surface
can be adequately described  by the  vacuum/solid interface within a
unit cell  which is sufficiently large along a certain direction. By utilizing the
so-called decimation technique~\cite{zhang2009BiSb} such a unit cell can be extended  to
infinity  which ensures  correct boundary conditions. Such calculations
typically require a formalism using fast decaying basis functions. In
the present work we apply the Wannier functions approach~\cite{souza2001}. As it follows from the band structures  shown in
Figure~\ref{FIG:surface-states} only  LiAuSe exhibits  gapless surface
states seen as a single Dirac cone within the bulk energy gap at the
$\Gamma$ point. The estimate of its Fermi velocity gives about
$1.8\times10^5\,$\,m/s which is smaller than that for Bi$_2$Se$_3$.
In contrast, KZnP and KHgSb remain insulators at the surface.  Thus the surface state
calculation agrees  with the bulk parity analysis and conclusively confirms
the existence of the topologically non-trivial  materials within proposed honeycomb type structure.

In conclusion we emphasize that topologically non-trivial systems
can be found in the proposed class of honeycomb structure
semiconductors. The interplay of mechanisms which are responsible for the
topologically trivial or non-trivial character in these systems differs from the
 cubic semiconductors studied earlier. In particular it is shown that
 the strong SOC and weak inter-layer coupling causes a double inversion
 which in turn makes the compound trivial. In
 contrast to the topologically non-trivial cubic systems which exhibit
 a zero band gap in the bulk, the topologically non-trivial hexagonal  materials
 provide the ``natural'' 3D topological materials with a real bulk band
 gap and gapless states at the surface.

\begin{acknowledgments}
The work was supported by the supercomputing center at Stanford
Institute Materials and Energy Science. The financial support of the
DFG/ASPIMATT project (unit 1.2-A) is gratefully acknowledged.
\end{acknowledgments}

\end{document}


\title{Supplementary material: Topological Insulators in Ternary Compounds with a Honeycomb Lattice}

\author{Hai-Jun Zhang$^{1}$,  Stanislav Chadov$^{2}$, Lukas M\"{u}chler$^{2}$, Binghai Yan$^{1}$, Xiao-Liang Qi$^{1}$, J\"{u}rgen K\"{u}bler$^{3}$, Shou-Cheng Zhang$^{1}$, Claudia Felser$^{1,2}$}

\affiliation{
  $^1$Department of Physics, McCullough Building, Stanford University,
  Stanford, California 94305-404531\\
  $^2$Institut f\"ur Anorganische Chemie und Analytische
  Chemie, Johannes Gutenberg - Universtit\"{a}t,  55099 Mainz, Germany,\\
  $^3$Institut f\"{u}r Festk\"{o}rperphysik, Technische Universit\"{a}t Darmstadt, 64289 Darmstadt, Germany
}

\email{felser@uni-mainz.de}

\date{\today}



\begin{abstract}
This is supplementary material. We provide the details of electronic structure calculation, parity
definition and illustrate the analogy in the band structures of
cubic and hexagonal semiconductors.
\end{abstract}

\maketitle

Calculation of the electronic structure  is performed by the plane-wave based
pseudopotential BSTATE (Beijing Simulation Tool of Atomic TEchnology) 
package~\cite{fang2002}. The exchange-correlation potential is treated
within the Generalized Gradient Approximation (GGA)~\cite{PBE96}. The lattice parameters listed in
present work are obtained by doing the {\it ab-initio} geometry optimization.

The topological character is calculated by following Ref.~\cite{Fu07}. The central quantity here
is the parity product  ${\delta_i=\prod_{m=1}^{N}\xi_{2m}(\Gamma_i)}$, where $N$ is the number of filled bands and
$\xi_{2m}(\Gamma_i)$ is the parity eigenvalue of the 2$m$-th occupied band
at the time reversal point $\Gamma_i$. The $Z_2$ invariant, $\nu_0=0$ or 1, is then obtained
from the product $(-1)^{\nu_0}=\prod_{i}\delta_i$.

\begin{figure}[htb!]
\centering
  \includegraphics[angle=270,width=0.90\linewidth,clip]{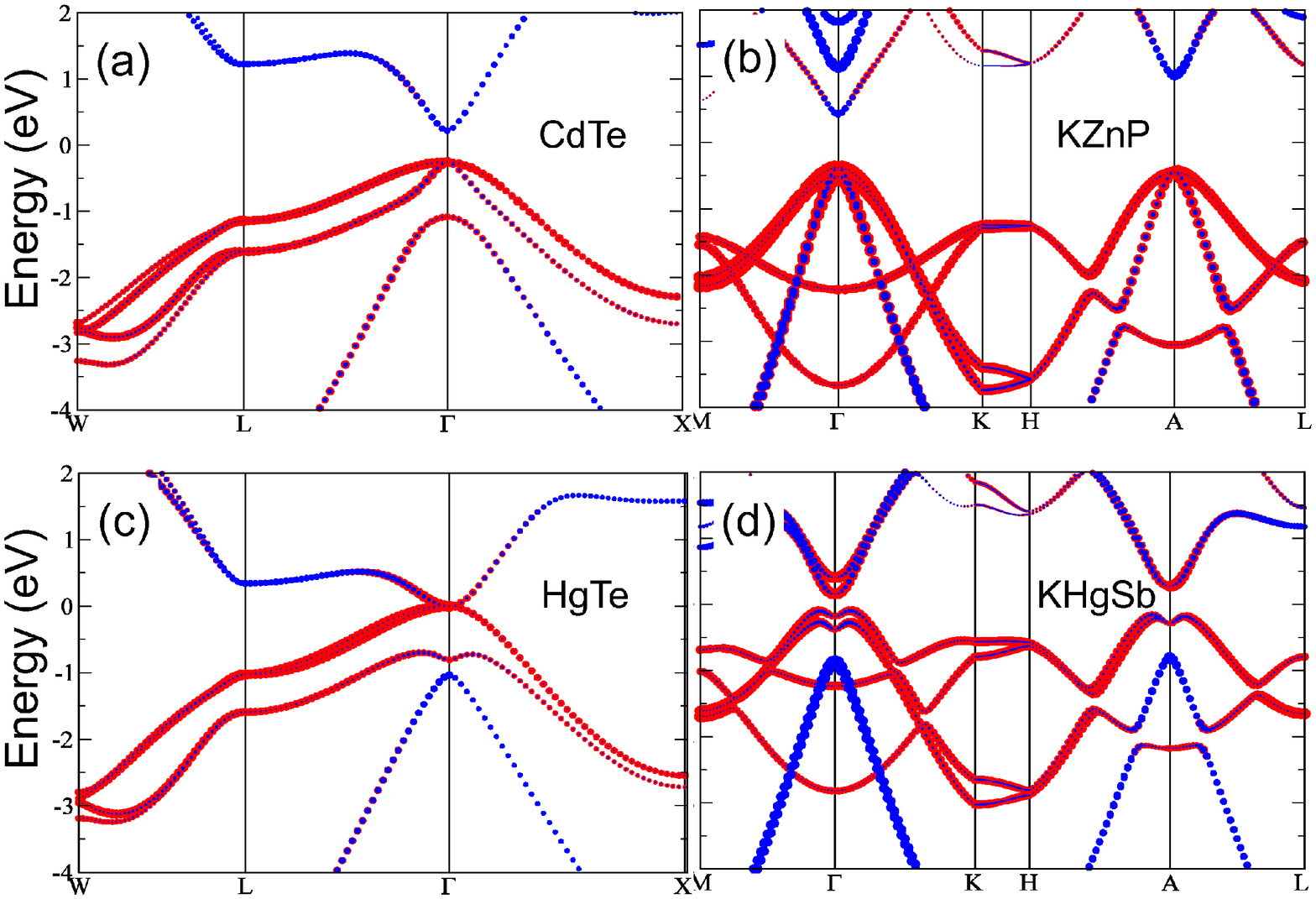}
\caption{(color online) Comparison of the band structures of cubic binaries
  (CdTe, HgTe) with hexagonal (KZnP, KHgSb) ternary compounds.
In both CdTe and KZnP the valence and conduction bands are formed
mainly by the $p$ (red) and the $s$ (blue) states,  respectively.}
\label{FIG:topological_fingerprint}
\end{figure}

Figure~\ref{FIG:topological_fingerprint} illustrates the analogy between
cubic C$_{1b}$ compounds and hexagonal semiconductors, by comparing the band structures of CdTe and
HgTe with hexagonal ternaries KZnP and KHgSb.
Clear fingerprints can immediately be identified: both CdTe and KZnP
exhibit a direct gap at the $\Gamma$ point formed by splitting
of the conduction ($s$-type) and the valence ($p$-type) bands. By going
from CdTe to HgTe and from KZnP to KHgP the bands at the $\Gamma$ point
invert in a similar manner: the $ s$-type  shifts below the $p$-type band.
However, in contrast to HgTe and other C$_{1b}$
Heuslers, the hexagonal symmetry with varying $c/a$ ratios  allows to lift the degeneracy
of the $p$-type bands and opens a finite band gap. As we mentioned this suggests
KHgSb and  the related materials as candidates for a 3D
topological insulators, similar to  Bi$_2$Se$_3$.  It  follows from the
flat bands along the $K-H$ direction that the material is quasi two-dimensional
which leads to a strong anisotropy of its transport properties.

%